\documentclass[twocolumn]{revtex4}

\usepackage{graphicx}%

\begin{document}

\title{Next stage of search for 2K(2$\nu$)-capture of $^{78}$Kr}

\author{Ju.M. Gavriljuk}
\author{V.N. Gavrin}
\author{A.M. Gangapshev}
\author{V.V. Kazalov}
\author{V.V. Kuzminov\footnote {\texttt{Talk at 5th International Conference on
Non-Accelerator New Physics (NANP-05), Dubna, Russia, 20-25 June
2005.}}}

\author{N.Ya. Osetrova}
\address{{\it
Baksan Neutrino Observatory, Institute for Nuclear Reseach, Russian
Academy of Sciences.}}

\author{I.I. Pul'nikov}
\author{A.V. Ryabukhin}
\author{A.N. Shubin}
\author{G.M. Skorynin}
\address{{\it
FSUE PA "Electrochemical Plant", Krasnoyarsk
Region,Russia}}

\author{S.I. Panasenko}
\author{S.S. Ratkevich}
\address{{\it
Karazin Kharkiv National University, Ukraine.}}


\begin{abstract}
A technique to search for 2K-capture of $^{78}$Kr with large
low-background proportional counter filled with an enriched in
$^{78}$Kr up to 99.8\% sample of Krypton at a pressure of 4.51 is
described in this paper. The results of first measurements are
presented. Analysis of data collected during 159 hours yielded new
limit to the half-life of $^{78}$Kr with regard to 2K-capture
(T$_{1/2}\geq6\cdot10^{21}$ yr ($90\%$ C.L.)). Sensitivity of the
facility to the process for one year of measurement was evaluated
to be $\texttt{S}=1.0\cdot10^{22}$ yr ($90\%$ C.L.).
\end{abstract}

\maketitle

\begin{center} {\bf Introduction} \end{center}

The up-to-date experimental limit to half-life of  $^{78}$Kr with
regard to 2K(2$\nu$)-capture is
$\texttt{T}_{1/2}\geq2.3\cdot10^{20}$ yr (90\% C.L.) \cite{ref1}.
Theoretical calculations based on various models yield the
following half-lives of  $^{78}$Kr with regard to double electron
capture of $^{78}$Kr(2e,2$\nu$)$^{78}$Se: $3.7\cdot10^{21}$ yr
\cite{ref2}, $3.7\cdot10^{22}$ yr \cite{ref3}, $6.2\cdot10^{23}$
yr \cite{ref4}. Fraction of events for 2K(2$\nu$)-capture for this
isotope constitutes 78.6\% of the total number of
2K(2$\nu$)-captures \cite{ref5}, thus yielding
$\texttt{T}_{1/2}$(2K,2$\nu$) to be $4.7\cdot10^{21}$ yr,
$4.7\cdot10^{22}$ yr, $7.9\cdot10^{23}$ yr, respectively. It is
clear from the comparison of the experimental and theoretical
limits that the sensitivity of the experiment must be improved by
20 or more times to verify the validity of the theoretical models.
The technique used to enhance sensitivity and the results achieved
are described in this paper.


\begin{center} {\bf Experimental technique and setup} \end{center}

The reaction $^{78}$Kr(2e,2$\nu$)$^{78}$Se resulted in the
production of $^{78}$Se$^{**}$ atom with two vacancies on K-shell.
The technique of searching for this reaction is based on the
assumption that energies of the characteristic photons and
probabilities of their emission in filling the double vacancy
coincide with the corresponding values proper for filling single
vacancy on K-shell for two separate singly ionized Se$^*$ atoms.
Then total registered energy is equal to 2K$_{ab}=25.3$ keV, where
2K$_{ab}$ - is the binding energy of K-electron in atom of
Se(12.65 keV). Fluorescence yield at filling the single vacancy
for K-shell of Se is equal to 0.596. Energies and relative
intensities of the characteristic lines of K-series are equal to
K$_{\alpha1}=11.22$keV (100\%), K$_{\alpha2}=11.18$ keV (52\%),
K$_{\beta1}=12.49$ keV (21\%), K$_{\beta2}=12.65$ keV (1\%)
\cite{ref6}. Probabilities of de-excitation with Auger electron
emission ($e_a,e_a$) only, or with a single characteristic quantum
and Auger electrons (K,$e_a$), or with two characteristic quanta
and low energy Auger electrons (K,K,$e_a$) are
$\texttt{p}_1=0.163$, $\texttt{p}_2=0.482$ and
$\texttt{p}_3=0.355$, respectively.

A characteristic quantum in a gas can pass a long enough distance
from a point of its origin to the point of absorption. 10\% of
characteristic quanta with energies of 11.2 keV and 12.5 keV are
absorbed in Krypton at a pressure of 4.35 atm ($\rho=0.0164$
$g/cm^{3}$) at a length of 1.83 mm and 2.42 mm, respectively (data
on absorption coefficient were taken from work \cite{ref7}. Runs
of photoelectrons with the same energies are 0.37 mm and 0.44 mm,
respectively. Their energy release is practically point-like. In
case of two characteristic X-rays emission and absorption in the
working gas the total energy will be distributed over three
point-like regions. It is just these events, which have a unique
set of parameters, that are the subject of research in the present
paper.

To register the above mentioned process we use a large
proportional counter (PC) with a body made of copper of M1 type.
The cylindrical body has the following inner dimensions: diameter
of 140 mm, and length of 710 mm. The anode wire made of
gold-plated tungsten with a diameter of 10 $\mu$m is stretched
along the cylinder axis. The total volume of the counter is 10,4
l, its working one is 9.16 l. PC is surrounded by the
low-background shield consisting of borated polyethylene (8 $cm$)
+ lead (15 $cm$)
 + copper (18 $cm$). The facility is located in the separate room
 of the underground laboratory of the Gallium-Germanium Neutrino
 Telescope of the Baksan Neutrino Observatory INR RAS at a depth
 of 4700 w.m.e. A sample of Kr of 47.65 l volume enriched in $^{78}$Kr
 up to 99.81\% is used to fill the PC. It was produced at FSUE
 PA "Electrochemical plant", Zelenogorsk from a Krypton sample
 used in work \cite{ref1}.  Content of $^{78}$Kr in the original sample was 94\%.
 There was an admixture of radioactive $^{85}$Kr with volume activity
 of  0.14 Bq/l which produced a background in the PC.
 The main purpose of reprocessing of this sample was to lower content
 of $^{85}$Kr. Finally $^{85}$Kr activity in the new sample was
 decreased by at least 55 times \cite{ref9}. Before filling the PC the gas
 was purified in the reactor with Ni/SiO$_2$  getter.
 Krypton pressure in the PC is 4.51 at. High voltage of 2400 V is applied
 to the anode. A signal is taken by charge-sensitive preamplifier
 (CSP) through high-voltage coupling capacitor. Signals, after
 their amplification in the additional amplifier, go to the input
 of the digital oscilloscope LA-n20-12PCI, which is inserted into
 the personal computer. The oscilloscope reads them with the
 digitized frequency of  6.25 MHz. This technique of pulse
 registration allows one to exclude the procedure of taking
 signals from both ends of the anode wire, as was the case
 in work \cite{ref1} where this procedure was used to determine the
 coordinates of the event and eliminate pulses produced by
 microdischarges outside the working length of the PC from
 its amplitude spectra.

In the newly devised layout the same result was achieved by
analyzing the signal shape. Pulses of microdischarges have
significantly faster rise-time compared with pulses from
ionization. Calibration of the PC was carried out through the wall
of the copper body (h=5 mm) by gammas of the source of $^{109}$Cd
(88 keV). In Fig. 1 one can see one of the pulses of the source
($a$), and its smoothed differential ($b$) for the case of 88 keV
quanta absorption on the Kr-atom K-shell resulting in emission of
a photoelectron and characteristic photon K$_{Kr}$ with energy
12.6 keV (two-point event).
\begin{figure}
\includegraphics*[width=5.5cm,angle=270.]{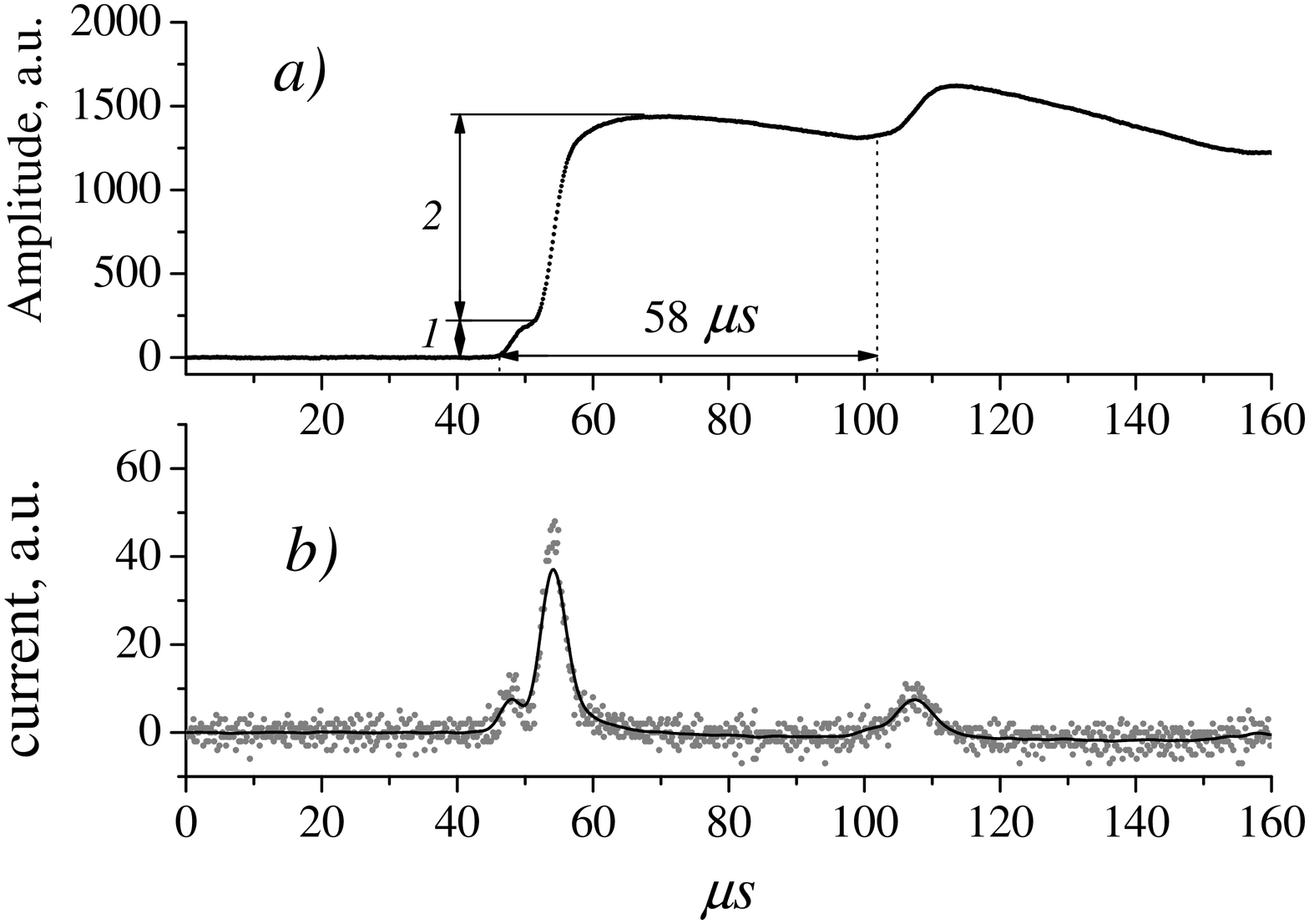}
\caption{One of the pulses of the source ($a$), and its smoothed
differential ($b$) for the case of 88 keV quanta absorption on the
Kr-atom K-shell resulting in emission of a photoelectron and
characteristic photon K$_{Kr}$ with energy 12.6 keV (two-point
event).}
\end{figure}
Fluorescence yield for K-shell of Krypton is 0.660 \cite{ref6}.
Maximum distance between point-like energy releases in the
projection to the radius of PC is equal to the radius. Drift time
for the ionization electrons moving from the cathode to the anode is
58 $\mu s$.  It is clear from Fig. 1($a$) that after ~58 $\mu s$
there is the second pulse in the PC whose amplitude is by ~8 times
lesser than that of the first. It is produced by secondary
photoelectrons knocked out from the cathode by photons produced in
the avalanche development process caused by primary ionization. The
presence of photo effect at the cathode is very probable due to the
absence of quenching admixtures in the working gas.

In general, comparison of the amplitudes and parameters of the
primary and secondary pulse shapes, allows one to exclude from the
spectrum the background events recorded near end-wall insulators
which support the anode.

To minimize the end-effect, the part of the anode coming out of
the insulator was thickened with copper tubes at a length of 39
mm. It is possible to collect charge in this region in the
ionization mode without photon generation. Collection of charge of
primary ionization in the ionization or combined mode affects the
shape of the first pulse and amplitude of the second.
\begin{figure}
\includegraphics*[width=7.5cm,angle=0.]{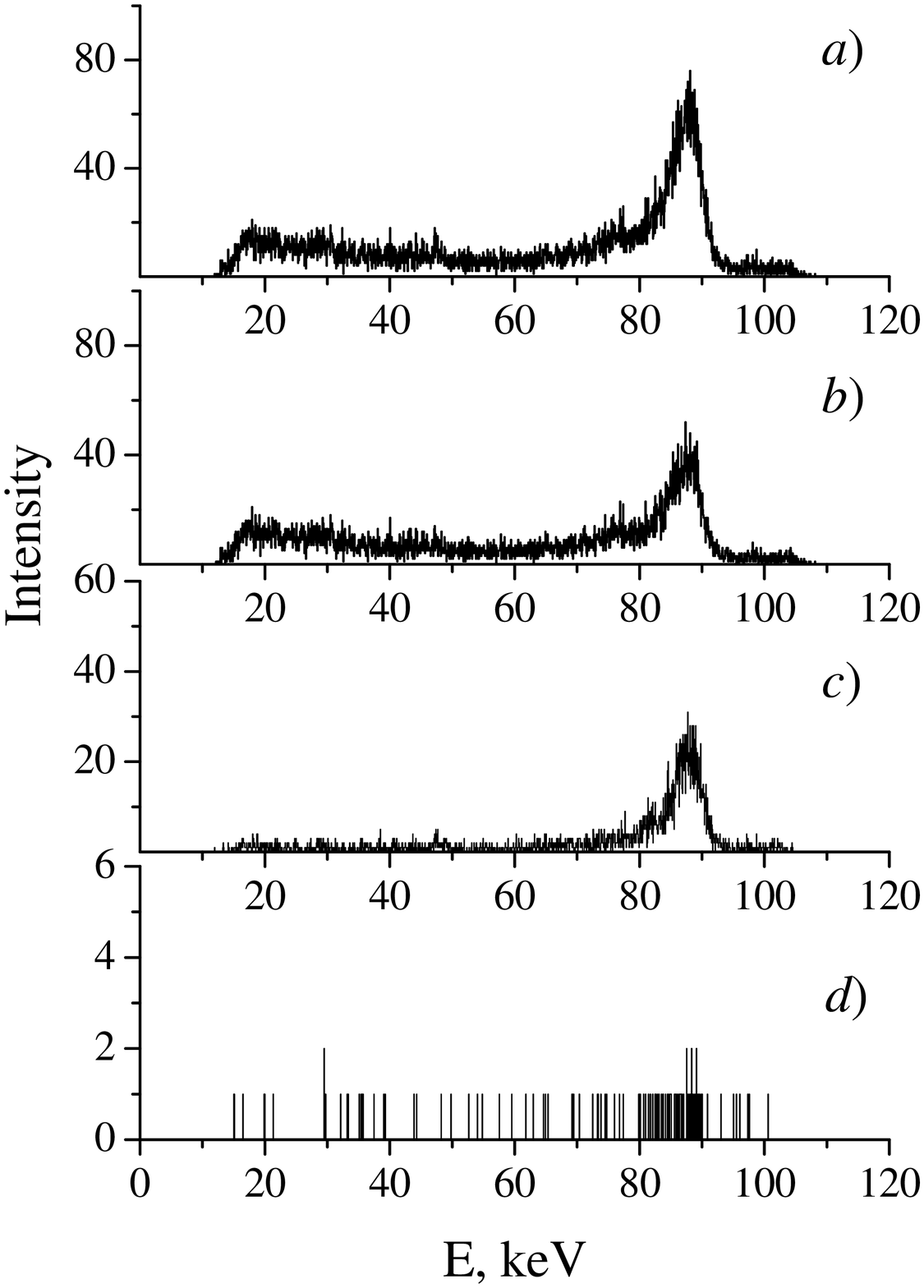}
\caption{The spectrum of 88 keV line ($^{109}$Cd) is given for all
events ($a$), for one-point events ($b$), for two-point events ($c$)
and for three-point events ($d$).}
\end{figure}
In Fig. 2 the spectrum of 88 keV line is given for all events ($a$),
for one-point events ($b$), for two-point events ($c$) and for
three-point events ($d$). Resolution of the 88 keV line in spectrum
($a$) determined for the right slope is 6\%. Ratio of peak 88 keV
squares in spectra ($b$) and ($c$) is determined by the following
factors: the fluorescence yield from K-shell of Kr, the efficiency
of absorption of characteristic emission of Kr in the working
volume, the ratio of photo effect probability on K- and L-shells,
the efficiency of selection of two-point events. In Fig. 3 spectra
for separate components of two-point events taken from Fig. 2($c$)
are plotted for the components with larger ($a$) and smaller ($b$)
amplitudes. Peak at Fig. 3$a$ corresponds to energy
\begin{eqnarray*}
(\texttt{E}_\gamma -\texttt{E}_{K\alpha Kr})=88 \texttt{keV}-12.6
\texttt{keV} =75.4 \texttt{keV},
\end{eqnarray*}
peak in Fig.3$b$ corresponds to $\texttt{E}_{K\alpha Kr}$.
\begin{figure}
\includegraphics*[width=6.5cm,angle=0.]{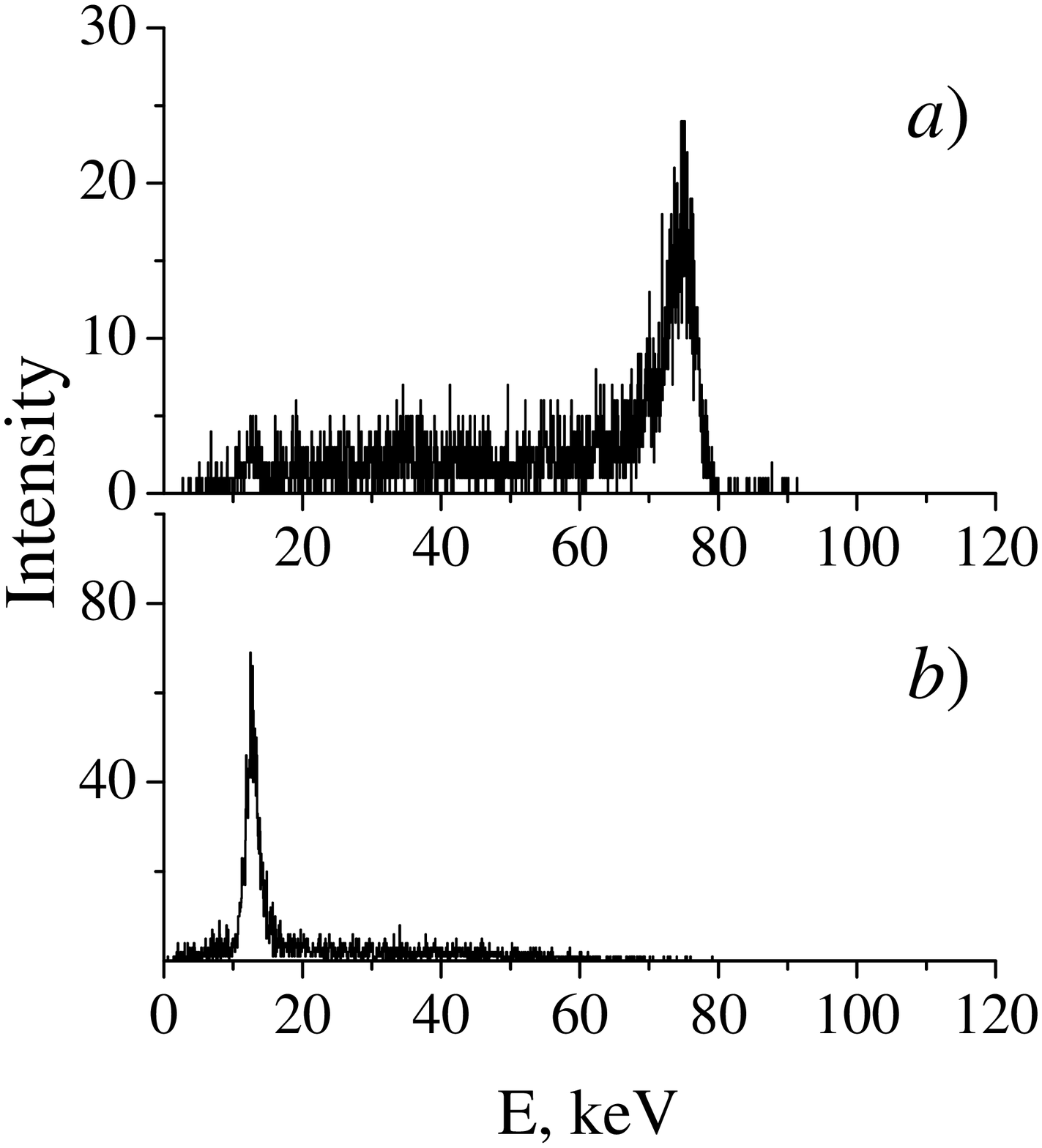}
\caption{The spectra for separate components of two-point events
taken from Fig. 2($c$) are plotted for the components with larger
($a$) and smaller ($b$) amplitudes.}
\end{figure}
Resolution of these peaks is 6.5\% and 15.9\%, respectively. In Fig.
4 one can see spectra of PC background collected for 159 hours for
all events ($a$), for one-point events ($b$), for two-point events
($c$) and three-point events ($d$). There is a peak of $\gamma$-line
of $^{210}$Pb seen in the spectra 4($a-d$) at the energy of 46.5
keV. It is supposed to be the daughter isotope of Rn-222 deposited
on the outer surface of the PC body during storage time in the
underground laboratory. Integrated over the spectrum PC count rate
versus time is shown in Fig. 5. As is seen, there is a
non-equilibrium Rn-222 brought with a gas from the Ni/SiO$_2$
reactor. There is also approximation by the dependence
\begin{eqnarray*}
\texttt{F}=49\cdot\exp(-\frac{0.693}{3.82\cdot 24}\cdot t)+42.
\end{eqnarray*}

The expected mean count rate after non-equilibrium Radon decay is
42 h$^{-1}$.

\begin{figure}
\includegraphics[width=6.5cm,angle=0.]{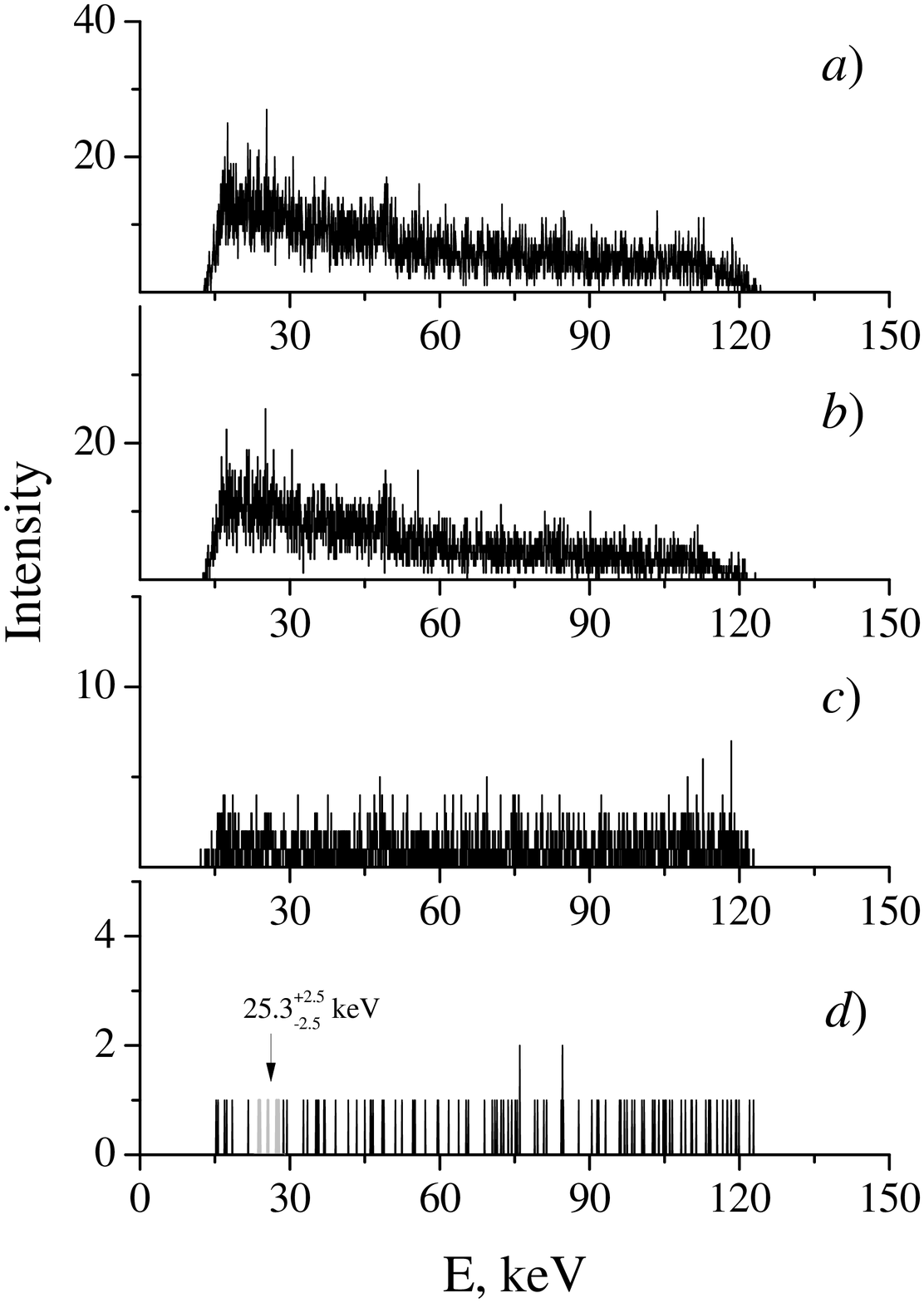}
\caption{The spectra of PC background collected for 159 hours for
all events ($a$), for one-point events ($b$), for two-point events
($c$) and three-point events ($d$).}
\end{figure}
\begin{figure}
\includegraphics*[width=5.5cm,angle=-90.]{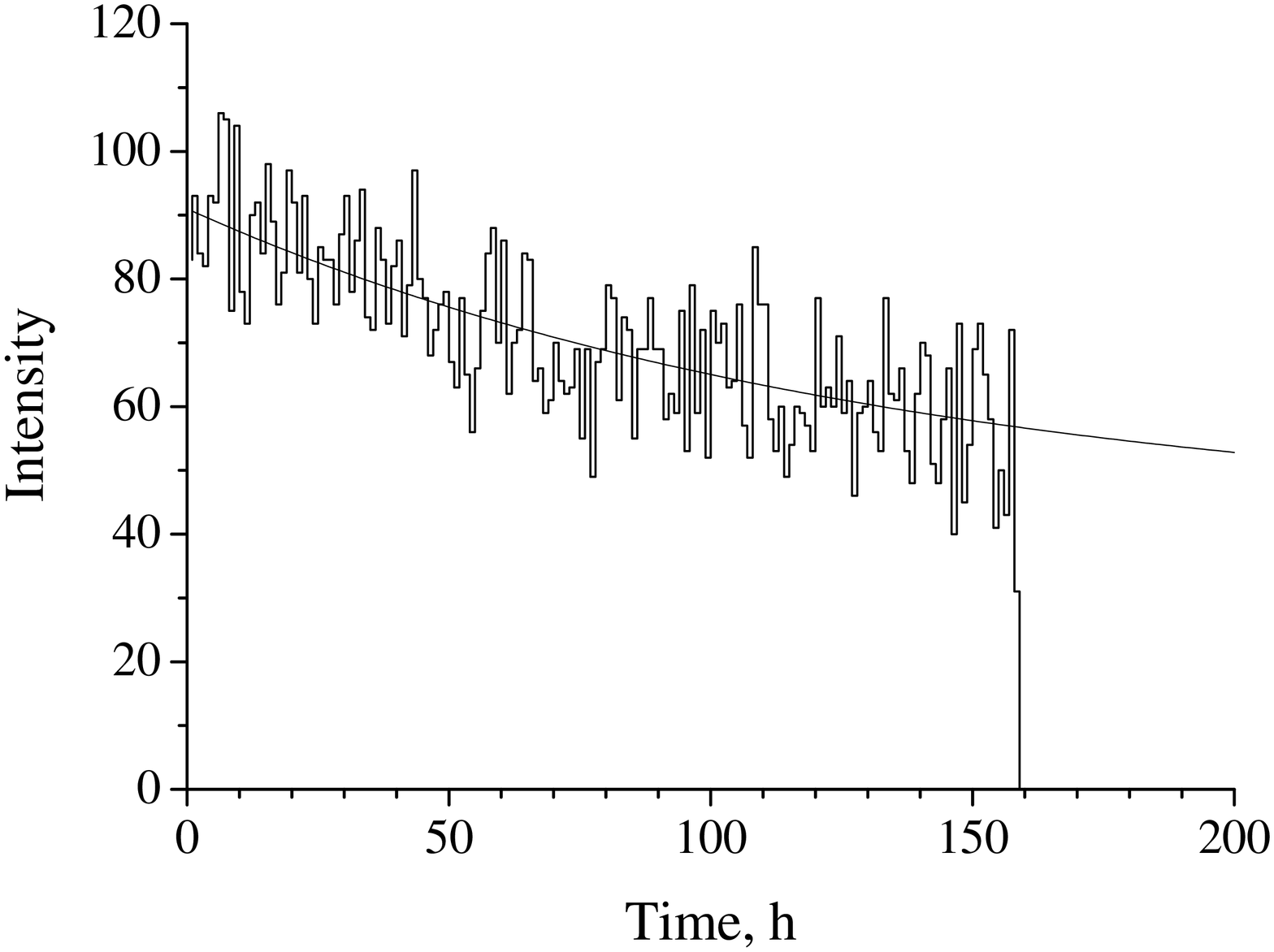}
\caption{Integrated over the spectrum PC count rate versus time.}
\end{figure}
To search for the possible effect in the spectrum 4 ($d$) the region
of $25.3\pm2.5$ keV was used. 95\% of useful events are coming in
there. For each event amplitudes of all three components are
compared with the expected calculated values for whom intervals of
the change are taken in the range of  $\pm2\sigma$ (95\% events).
Parameter  is determined according to the individual energetic
resolution ($\texttt{R}$) of the expected lines
($\texttt{R}=2.36\cdot \sigma$). The analysis starts with a
comparison of the component of the minimum energy. For the case
($\texttt{K}_{\alpha},\texttt{K}_{\alpha},\texttt{e}_{\texttt{a}}$)
 the energy E$_{\texttt{e}_\texttt{a}}$ is $25.3-2\cdot11.2=2.9$ keV
 (probability of the even is $\texttt{p}_4=0.773$), for the case
 ($\texttt{K}_{\alpha},\texttt{K}_{\beta},\texttt{e}_{\texttt{a}}$)
 the energy E$_{\texttt{e}_\texttt{a}}$ is 1.6 keV ($\texttt{p}_5=0.213$).
 Then goes a comparison of two other components with the calculated values.
 There was no event with full set of the calculated features among
 those five that came into the working energy range.
 Thus we can conclude that the effect is zero. According to
 \cite{ref10}
 the possible effect ($\texttt{M}$) (at 90\% C.L.) does not exceed
 2.44 for the period of measurement or $\texttt{M}\leq 134.4$ year$^{-1}$.
 Calculation of the limit to the half-life is carried out with the formula:
\begin{eqnarray*}
\texttt{T}_{1/2}\geq \frac{\ln2 \cdot N \cdot \varepsilon_1 \cdot
\varepsilon_2 \cdot (\texttt{p}_4 + \texttt{p}_5) \cdot 0.95}
{\texttt{M}},
\end{eqnarray*}
$N=1.08\cdot 10^{24}$ is the number of $^{78}$Kr atoms in the
working volume, $\varepsilon_1=0.809$ is the efficiency of
registration of the two-photon events, $\varepsilon_2\approx 1$ is
the efficiency of selection of the three-point events, and
$\texttt{p}_3-\texttt{p}_5$ are described above. Thus we find:
\begin{eqnarray*}
\texttt{T}_{1/2}(2\texttt{K},2\nu+0\nu)\geq 1.5\cdot10^{21} \,\,
\texttt{yr} \,\,(90\% \,\,\texttt{C.L.}).
\end{eqnarray*}

Sensitivity of the setup to the sought process could be calculated
if one assumes that there is no effect and value \texttt{M}
determined above is a background. Thus for one year of measurement
one can find an excess over background to be $1.645\sigma$ (90\%
C.L.), where  $\sigma=\sqrt{\texttt{M}}$, or
\begin{eqnarray*}
\texttt{S}=1\cdot10^{22}\,\,\texttt{yr}\,\,(90\% \,\,\texttt{C.L.}).
\end{eqnarray*}
Measurement is in process.

The work was done under the financial support of the RFBR (grant
no. 04-02-16037) and "Neutrino Physics" Program of the Presidium
of RAS.

\end{document}